\documentclass[pdflatex,sn-mathphys]{sn-jnl}% Math and Physical Sciences Reference Style
\jyear{2021}%

\theoremstyle{thmstyleone}%
\theoremstyle{thmstyletwo}%

\theoremstyle{thmstylethree}%

\begin{document}

\title[Two-stage dynamic creative optimization ]{Two-stage dynamic creative optimization under sparse ambiguous samples for e-commerce advertising}

\author*[1]{\fnm{Guandong} \sur{Li}}\email{leeguandon@gmail.com}
\author[1]{\fnm{Xian} \sur{Yang}}

\affil*[1]{\orgdiv{AI Center}, \orgname{Suning}, \orgaddress{\street{Xuanwu}, \city{Nanjing}, \postcode{210042}, \state{Jiangsu}, \country{China}}}


\abstract{Ad creative is one of the main mediums for e-commerce advertising. Ad creative with good visuals may increase a product's click-through rate(ctr). In recent years, unlike artificially produced ad creatives, advertising platforms can automatically synthesize ad creatives, and each type of element can be arbitrarily specified. Advertisers only need to provide basic materials to synthesize a large number of potential ad creatives in batches. But with limited real-time feedback, it is difficult to accurately estimate the ctr of creatives. In our scenario, in addition to a large number of sparse samples, we also face the problem of ambiguous samples. In our approach we decouple this dynamic creative optimization into two stages, a cascaded structure that can trade off between effectiveness and efficiency. In the first stage, we train an automatic creative optimization architecture based on autoco to simulate complex interactions between creative elements. Although we obtained the ranking of different creatives under a sku, because we bucketed and merged historical data according to periods, this confuses the ctr diversity of the same ad creatives on different days and weakens the ability to separate ambiguous samples. Therefore, we propose a transformer-based rerank model. With the help of the rank model, we propose a distillation method to learn the relative order of ideas and extract the ranking knowledge to guide the rerank learning. The creative order soft labels under each sku are generated by the rank model to alleviate the dilemma that a large number of under-represented creatives cannot obtain real labels. Through the knowledge diffusion of rerank, the ambiguous samples are associated with the positive and negative samples. Cascade rerank and autoco to output the estimated value of the synthetic ad image. In the second stage, we designed a bandit model, and the bandit selected one of the output ad of the first stage for timely delivery. Experimental results show that our method can outperform competing baselines in terms of sctr. Online A/B testing shows that our method improves ctr by 10\% compared to the baseline.}

\keywords{dynamic creative optimization, rerank, autoco, bandit,distillation,sparse and ambiguous samples}

\maketitle

\section{Introduction}\label{sec1}

Online display advertising is a fast-growing business and has become an important source of revenue for Internet service providers. Image ad is the most widely used form because it is intuitive, effective, and easy to understand\cite{bib1}. Traditionally, advertisers had to hire professional designers to create attractive creatives and then submit them to advertising platforms. Every time an advertiser launches a new product or updates an old one, they need to design many different sizes and styles of creative images. In recent years, advertising platforms can synthesize ad creatives in real time, and advertisers only need to provide basic materials to the platform, such as product images and text. According to these materials, the synthesis system synthesizes the ad creative composed of any specified elements such as template, text color, picture size, font, etc. as shown in Fig1. Although ad creatives represent the same product, their click-through rates can vary widely due to differences in visual appearance. Therefore, it is critical to present the most compelling ad creative to reach potentially interested consumers and maximize ctr. Compared with using artificial experience to select creatives, the inherent advantage of machine intelligence is that it can learn the pros and cons of creatives through massive data, and accumulate knowledge online, so as to select the ad creatives that can most attract consumers' attention.

Dynamic creative optimization: What the advertising platform receives is no longer a complete creative produced by hand, but various creative materials (such as template sets, image elements, points of interest, dimensions, etc.). The algorithm will dynamically adjust the selection of elements and optimize the production parameters according to user behavior information and delivery results, so that the final creative can take into account both visual effects and delivery effects. In the image synthesis system, many ad creatives are produced, but not all ad creatives have been exposed, so it is necessary to find the most suitable map for placement. For full creative granularity optimization, we can model it as a standard E\&E problem. Given a product, each candidate creative should have the opportunity to be shown to the user and get corresponding feedback (Explore stage). At the same time, in order to ensure the long-term benefits of advertising, the system will allocate more traffic to the creative with the best performance so far (Exploitation stage). The trade-off strategy between exploration and exploitation (E\&E) described above can be solved by the bandit model. Bandit models commonly used in business scenarios include Smoothed Epsilon-Greedy\cite{bib2}, Thompson Sampling\cite{bib3} and LinUCB \cite{bib4}.

In this paper, we will model our scene for dynamic creative optimization, and we face the following four problems:

1. Not all ad creatives have ctr, and not all images output by the synthesis system will be exposed. On the contrary, only the next several images of the same sku will be taken for delivery. Given the large number of potential ad creatives, with limited real-time feedback, ctr estimates often have high variance.

2. Our creative optimization is based on the synthetic image scene, the input is image, text, etc, we optimize the output image. Synthetic elements can be controlled, and the synthetic image nodes can be fused to impose constraints.

3. There are only a few ctr comparisons of multiple different creative images corresponding to one sku in the original delivery data. The output image of the same sku often selects only one image, which challenges us to construct the prior information of ad creative.

4. The lack of user-side features leads to repetition, large-scale promotions, targeted traffic to product pools, and completely different ctr values for the same creative on different days.

Faced with the above four problems, in our method, we decouple this dynamic creative optimization into two stages, and the cascaded structure can trade off between effectiveness and efficiency. In the first stage, the dynamic creative optimization system can synthesize images under the constraints of the input products and text, and the dynamic screening between elements is limited to a limited creative space. Composite elements include merchandise, copy, template groups, templates, dimensions. We train an automatic creative optimization architecture based on autoco\cite{bib5} to simulate complex interactions between creative elements. FM models interactions between creative elements based on inner products, so that ideas with similar composite elements are represented similarly. However, the interaction between ideas is more complicated. Only using the conventional inner product interaction function will make the model get suboptimal results. Therefore, the multiplication operator in the inner product is extended to the operator set {concat,multiply, plus,max,min}. After autoco, although we obtained the sorting of different creatives under a sku, due to problem 4, we bucketed and merged historical data periods, which confuses the ctr diversity of the same ad creatives on different days. The training of autoco is based on periodic bucketing and merging data. It has a good inference effect on relatively positive and negative samples, but it weakens the ability to separate ambiguous samples, and the training data is not based on the paired data of different ad creatives under the same sku. Therefore, we not only need to associate ambiguous samples, that is, the information of those samples that are not easily exposed, but also need to strengthen the information of mutual ordering under the same sku. Therefore, we propose a transformer-based rerank model\cite{bib6}, and with the help of the rank model\cite{bib7}, a distillation method\cite{bib8} is proposed to learn the relative order of ideas and extract the ranking knowledge to guide the rerank learning. The soft labels in each ad creative are generated by the rank model to alleviate the dilemma that a large number of under-displayed creatives cannot obtain real labels. Therefore, based on the first stage, we output five images based on the same sku. In the second stage, we designed a bandit model, and bandit selected one of the five pictures for timely delivery.

	Our contributions are:
	
1. We propose a two-stage dynamic creative optimization framework for optimally selecting ad creatives with composite elements. In the first stage, n choose 5, and in the second stage, choose 1 from 5. The framework considers the modeling of complex interactions between creative elements, and based on a large number of feedback sparse problems and ambiguous samples, a method for inter-order modeling is proposed by distillation.

2. In the first stage of the dynamic creative optimization framework, we propose a distillation rerank model under the rank model, and the transformer-based rerank model complements the shortcomings of creative element modeling.

3. We evaluate the proposed algorithm on synthetic datasets and real datasets, and the results show that the proposed method improves ctr by 10\% on the Suning outbound information flow scenario due to the baseline.

\begin{figure}[h]%
\centering
\includegraphics[width=0.9\textwidth]{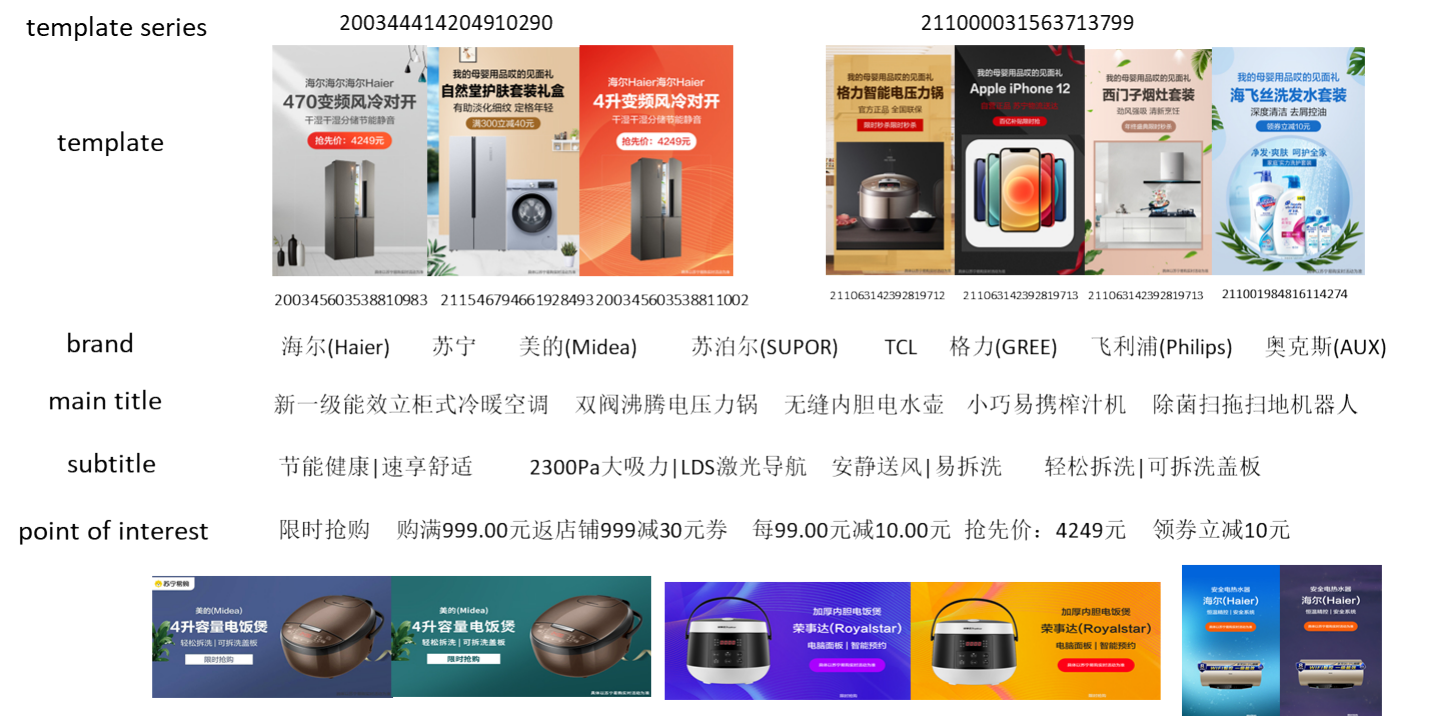}
\caption{\centering the combination of creative elements in the suning scene}\label{fig1}  
\end{figure}

\section{Related work}\label{sec2}
\subsection{problem formulation}\label{subsec2}

We consider a search advertising system with query q$\in$Q and ad set $[{a_{j}}]_{n}^{j=1}$, Each ad contains m creatives $a_{j}=[{c_{i}}]_{m}^{i=1}$,where the i-th creative $c_{i} =(v_{i},t_{i},id_{i},y_{i})$.Here $v_{i},t_{i},id_{i}$ and $y_{i}$  represent its image features, text features, creative ID features, and soft labels predicted by the ranking model, respectively. Our goal is to choose the best creative for each ad.

\subsection{dynamic creative optimization}\label{subsec2}
Dynamic creative optimization field, application scenarios include:1. for new scenarios, the data still has an initial stage with sufficient accumulation.2. In the long-tail part of mature scenarios, a large number of long-tail advertisements lack feedback data due to head effects and insufficient bids.3. Ads with frequent replacement/testing of creatives, advertisers change/add creatives to test the delivery effect. In these scenarios, Dynamic Creative Optimization can be used.

\cite{bib9} is given an ad schedule with hundreds of new ads. How to promote an ad with the highest potential profit among many creatives with minimal trial and error cost during the cold start phase. It proposes the Pre Evaluation of Ad Creative Model, a deep neural network model for the pre-tanking stage. It leverages images, taglines, OCR, and contextual features to evaluate content. The top-ranked ones are considered excellent ad creatives and will be fed into the online system and explored with priority. At the same time, it proposes a pairwise training method. Train a model to learn relative ranking instead of ctr. It turns creative optimization from ctr estimation into a ranking problem, and secondly, its sample labels are derived from human design.

A component tree-based adaptive efficient ad creative selection (AES) framework is proposed in \cite{bib10} to achieve optimal selection through dynamic programming. It also uses dynamic programming to adjust thompson sampling to efficiently explore the best ad creatives. It transforms the optimization problem of creative element parameter combination into the optimal path selection problem on the graph, and this problem can be solved quickly through the idea of dynamic programming. Due to the limited delivery data, adopting a tree structure can introduce visual prior information to narrow the search space.

A visual content-based creative effect prediction model was proposed in \cite{bib11}. The model is divided into Visual-aware Ranking Model (VAM) and Hybrid Bandit Model (HBM). The former learns the visual features related to the effect from the fully delivered data, while the latter uses the learned visual features and model parameters as the prior, and updates the posterior based on the actual delivery data.

An Automatic Creative Optimization (AutoCO) framework is proposed in \cite{bib5} to model complex interactions between creative elements and strike a balance between exploration and exploitation. Specifically, inspired by automl, it proposes a oneshot search algorithm to search for efficient interaction functions between elements, and developes stochastic variational inference to estimate posterior distributions of parameters based on reparameterization techniques, applying Thompson Sampling to efficiently explore better ideas.

A novel creative prioritization cascade architecture is proposed in \cite{bib7}, built before the rank model, to jointly optimize creative selection and ranking between ads. It designs a classic two-tower structure and allows creative feature extraction in the creative optimization stage to be shared with the ranking stage, and proposes a soft-label list ranking distillation method to extract ranking knowledge from the ranking stage to guide CACS learning. On the other hand, we also design an adaptive dropout network to encourage the model to probabilistically ignore ID features and use content features to learn multimodal representations of creatives.

\subsection{multi-armed bandit model}\label{subsec2}
The multi-armed bandit (MAB) problem is a typical sequential decision process and is also regarded as an online decision problem\cite{bib12}. A wide range of application scenarios can be modeled as MAB problems, such as recommender systems\cite{bib13}, online advertising\cite{bib14}, and information retrieval\cite{bib15}. Epsilon-greedy\cite{bib2}, Thompson sampling\cite{bib3} and UCB\cite{bib4} are classic context-free algorithms. They use the rewards/costs from the environment to update their E\&E policies without contextual information.

\subsection{learn2rank}\label{subsec2}
Learning to Rank (LTR)\cite{bib6} focuses on optimizing the global ranking of item lists according to the user's click-through rate on items. Applications in advertising scenarios are ubiquitous. When it comes to optimizing the order of search results, LTR provides a good result. It takes a dataset of search results and their click labels for users. Tags can be expressed as hierarchical relevance (artificial criteria such as 1 - 5) or binary relevance (usually a real-time search engine's click log). Given such a dataset, a ranking function can be learned to produce the desired order of items.

\section{Dataset}\label{sec3}
In our off-site dpa information flow delivery scenario, it is difficult for us to find paired data for the same sku. We compared the creative ranking dataset\cite{bib11}, the same ad creative of our same sku will be repeatedly placed and exposed in different days. Different from creative ranking, which can collect the ctr information of the same sku in the same time dimension (random sampling strategy and life cycle alignment), this is a difficult point in our scene data, and it is also the fundamental starting point of our one-stage design.As shown in Fig2.

\begin{figure}[h]%
\centering
\includegraphics[width=0.9\textwidth]{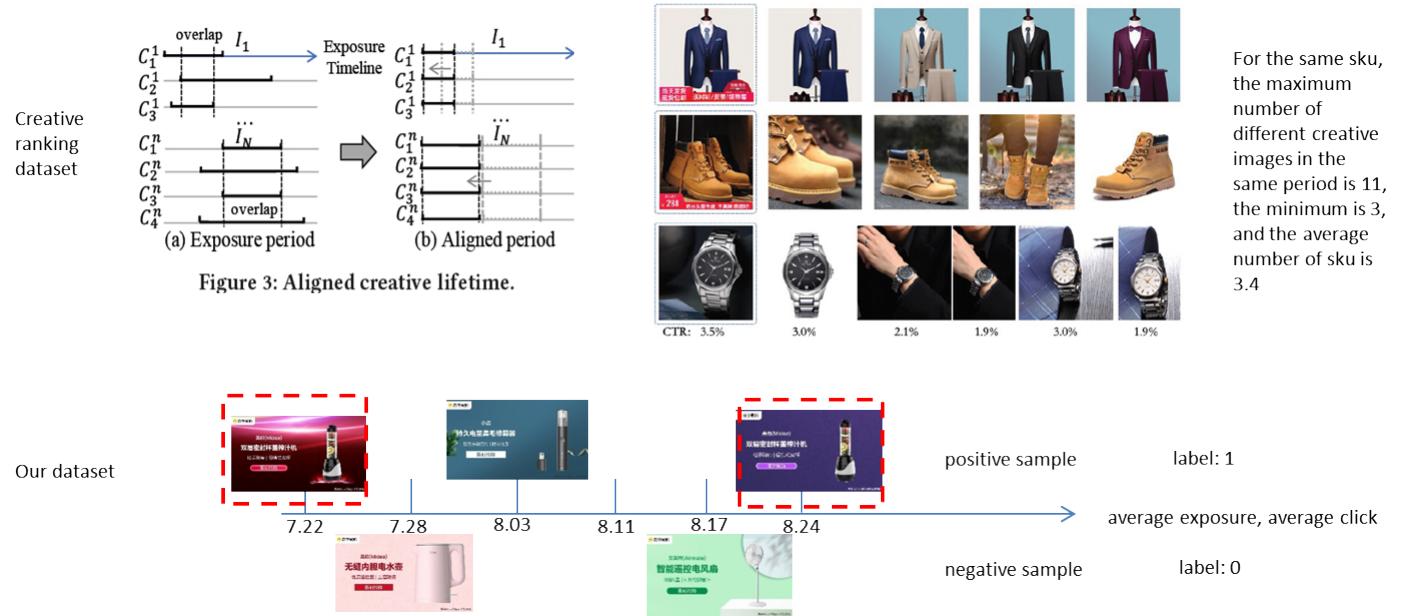}
\caption{\centering comparison of our scene data and the creative ranking dataset}\label{fig2}
\end{figure}

Here we illustrate several concepts that confront the data in our scenario:

Ambiguous sample: Some well exposed samples will show different ctr data on different days.

Sparse samples: Exposure to extremely small data, in principle, due to the sparsity of these samples, will not only lead to high variance of the prediction results, but also lead to the overall deterioration of the results.

Template series: Our set of elements includes commodity, text, template series, templates, and dimensions, where template series refers to a series of templates bound to the business department, and template is the template selected under the template series. For example, an refrigerator washing machine might correspond to 200344414204910290/211647091263912517/200344443301310314/211000031563713799.

We need to make positive and negative samples based on ctr, but we will encounter a large number of ambiguous samples and sparse samples, so how to make training data for autoco?

In order to exclude the influence of sparse samples, we select samples with exposure greater than average exposure and clicks greater than average clicks as positive samples, and samples with exposure greater than average exposure and clicks less than average clicks as negative samples. At this point, we think that the influence of sparse samples has been roughly ruled out. Let's briefly analyze the performance of positive and negative samples?

a. Template series distribution and business division distribution of positive and negative samples

\begin{table}[h]
\begin{center}
\begin{minipage}{\textwidth}
\caption{\centering template series exposure clicks ctr distribution}\label{tab1}%
\resizebox{\textwidth}{25mm}{
\begin{tabular}{@{}lccclccc@{}}
\toprule
& \multicolumn{2}{c}{negative sample} & \multicolumn{5}{c}{positive sample} \\
\cmidrule{1-4}\cmidrule{5-8}% 
template\_series\_id & exposure & clicks & ctr(\%) & template\_series\_id & exposure & clicks & ctr(\%) \\
\midrule
211647091263912517  & 1073515 & 3096 & 0.288 & 211647091263912517 & 8560166 & 157617 & 1.841 \\
200344440542910326  & 884838 & 2438 & 0.276 & 200344428100010286 & 6862785 & 117456 & 1.711 \\
200344443301310314  & 667765 & 1777 & 0.266 & 200344440542910326 & 6413013 & 80457 & 1.255 \\
200344428100010286  & 565730 & 1603 & 0.283 & 200344441482710312 & 5577912 & 85014 & 1.524 \\
200639166710918378  & 499090 & 1456 & 0.292 & 200639166710918378 & 3968730 & 93294 & 2.351 \\
211000031563713799  & 451218 & 1195 & 0.265 & 200344424245610379 & 3644707 & 81977 & 2.249 \\
200344424245610379  & 395128 & 1164 & 0.295 & 200344443301310314 & 3346416 & 51996 & 1.553 \\
200344441482710312  & 347736 & 1062 & 0.305 & 200344414204910290 & 3055797 & 75004 & 2.455 \\
200344433328410407  & 279896 & 741 & 0.265 & 211000031563713799 & 2830976 & 58459 & 2.065 \\
200344414204910290  & 252792 & 796 & 0.315 & 200344416496310282 & 2304206 & 48636 & 2.111 \\
200344416496310282  & 237374 & 700 & 0.295 & 200344432379510405 & 1958267 & 37043 & 1.892 \\
200344425785310295  & 233703 & 623 & 0.267 & 200344437754710322 & 1820407 & 50327 & 2.765 \\
200344438645910324  & 161191 & 444 & 0.275 & 200344438645910324 & 1581274 & 35665 & 2.255 \\
200344432379510405  & 139137 & 492 & 0.354 & 200344425785310295 & 1281158 & 22072 & 1.723 \\
\botrule
\end{tabular}}
\end{minipage}
\end{center}
\end{table}

\begin{table}[h]
\begin{center}
\begin{minipage}{\textwidth}
\caption{\centering distribution of corresponding delivery data of business divisions}\label{tab2}%
\resizebox{\textwidth}{30mm}{
\begin{tabular}{@{}cccc@{}}
\toprule
& \multicolumn{0}{c}{negative sample} & \multicolumn{1}{c}{positive sample} \\
\cmidrule{1-2}\cmidrule{3-4}% 
business department & Number of samples & business department & Number of sample \\
\midrule
supermarket  & 18780 & supermarket & 16359 \\
home devices & 5742 & communication & 5319 \\
kitchen and Bath Supplies  & 5112 & kitchen and Bath Supplies & 5114 \\
communication & 4779 & home devices & 4145 \\
refrigerator washing machine  & 3360 & refrigerator washing machine & 3909 \\
decoration supplies & 3125 & decoration supplies & 2477 \\
baby products  & 3095 & audiovisual multimedia & 1977 \\
food & 2391 & food & 1788 \\
design & 2168 & baby products & 1677 \\
office & 2166 & office & 1566 \\
toiletries & 1680 & air conditioner & 1405 \\
home appliances & 1409 & design & 1392\\
department store & 1365 & toiletries & 1356 \\
office & 2166 & office & 1566 \\
toiletries & 1680 & air conditioner & 1405 \\
audiovisual multimedia & 1364 & home appliances & 1114\\
computer & 1193 & computer & 695 \\
drinks & 1075 & drinks & 522 \\
car & 1019 & department store & 516\\
\botrule
\end{tabular}}
\end{minipage}
\end{center}
\end{table}

It can be seen that the positive and negative samples are confused under different template series and business units, which shows that the ad creative under the same sku is difficult to have obvious trends in the dimensions of business units and template series.

b. Size and color distribution of positive and negative samples
\begin{figure}[h]%
\centering
\includegraphics[width=0.9\textwidth]{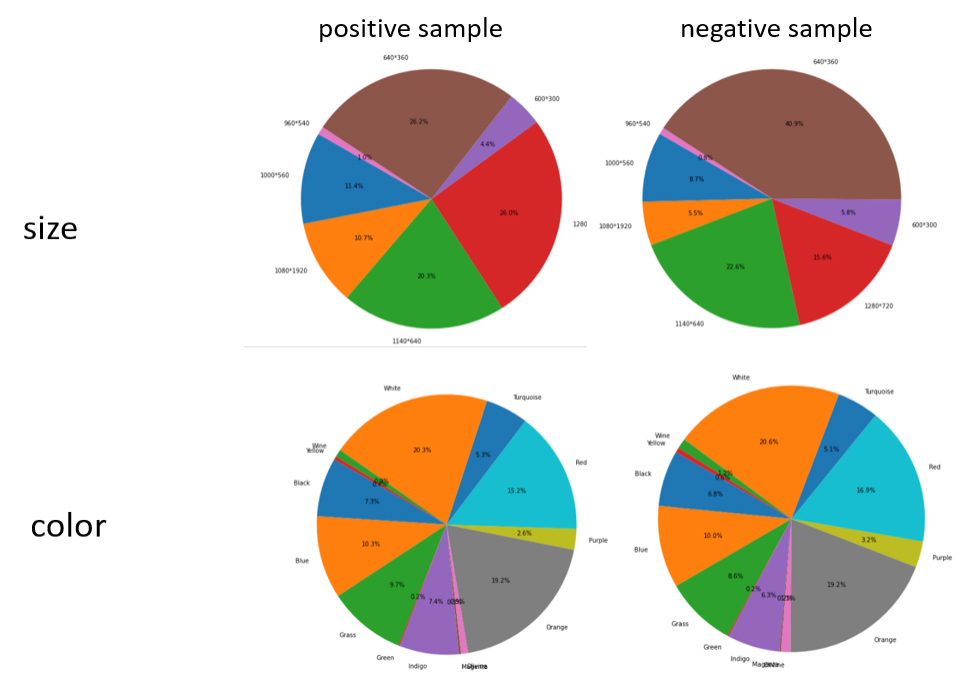}
\caption{\centering size and color distribution of positive and negative samples}\label{fig3}
\end{figure}
The color and size dimensions also hardly reflect the preferences of positive and negative samples.

c.Mixed positive and negative sample clustering, whether there is a cluster that is all positive or negative samples

\begin{figure}[h]%
\centering
\includegraphics[width=0.9\textwidth]{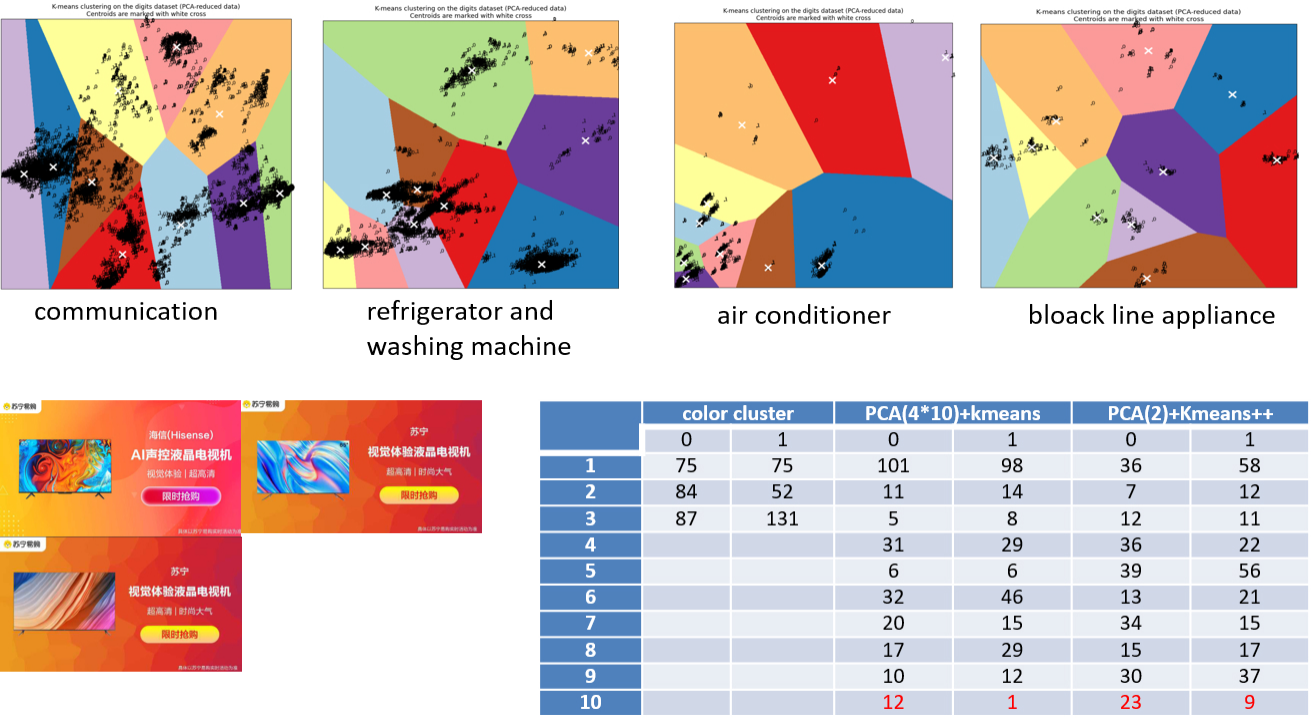}
\caption{\centering mixed positive and negative sample clustering}\label{fig4}
\end{figure}

Mixed positive and negative sample clustering, is there a cluster that is all positive samples or negative samples? So as to find out the characteristic difference in the distribution of positive and negative samples.The lower left is one of the clustered images. It can be seen that the colors are the same, and the clusters are still obvious.

d. Template series data analysis
\begin{figure}[h]%
\centering
\includegraphics[width=0.9\textwidth]{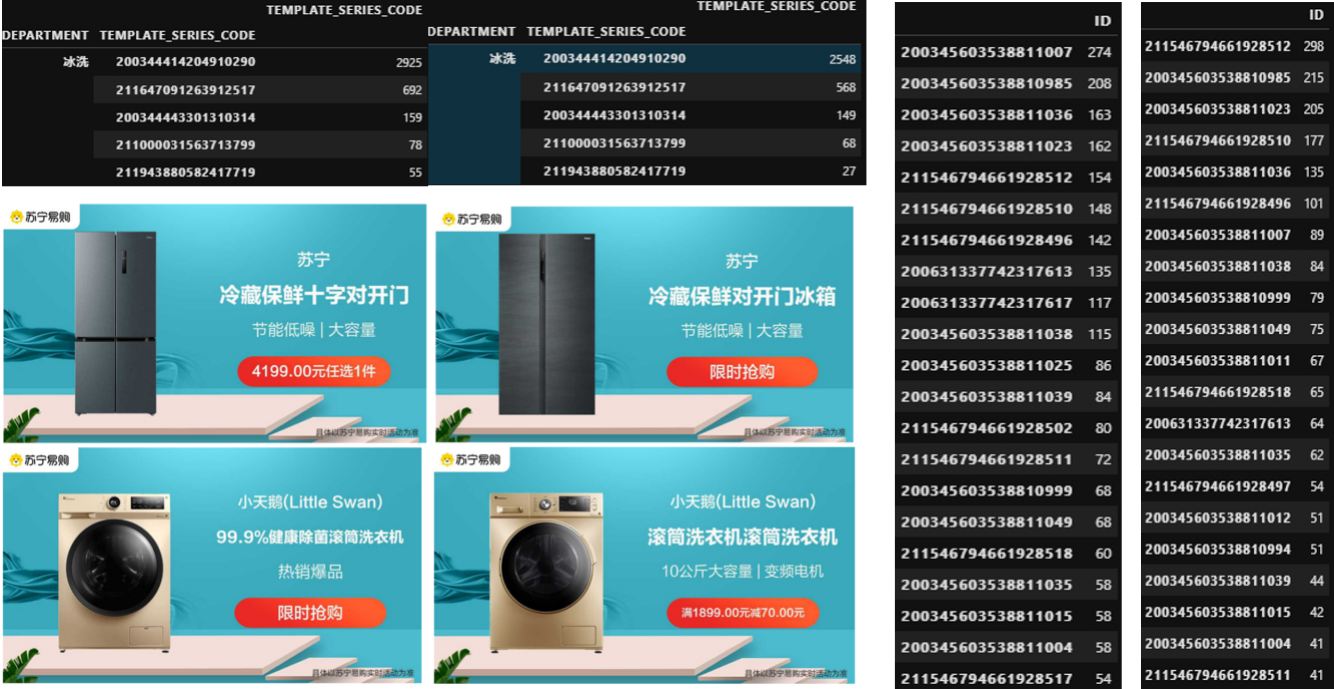}
\caption{\centering template series data analysis}\label{fig5}
\end{figure}

As shown in Fig5.Left: positive and negative samples under the same template.Right: distribution of positive and negative templates under the same template series. There is a one-to-many relationship between business units and template series. A business unit may correspond to multiple template series. Are there some templates under a template series that are relatively good factors? From the above, we find that there is no difference between different positive and negative sample ad creatives sets of the same sku.

Why does the above situation occur? We found that even some well-exposed samples show different exposure patterns on different days? We call it ambiguous samples,as shown in Fig6:

\begin{figure}[h]%
\centering
\includegraphics[width=0.9\textwidth]{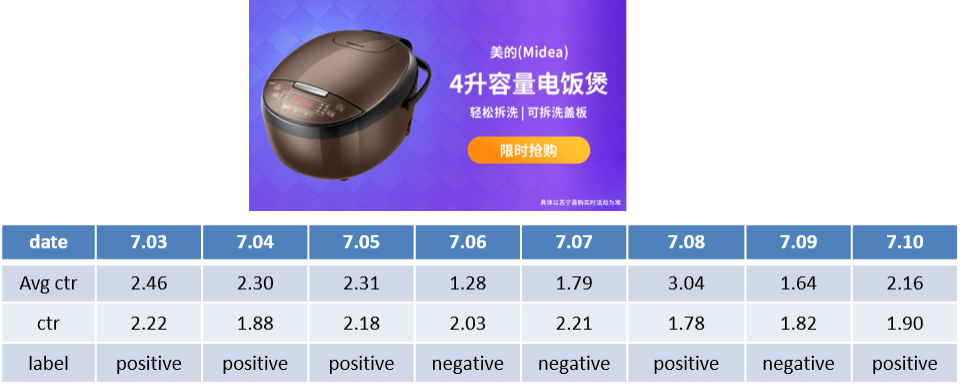}
\caption{\centering ambiguous sample example}\label{fig6}
\end{figure}

In the picture above, from July 3 to July 10, the same creative image may have different ctrs in different days due to continuous delivery, so there is ambiguity in the discrimination of positive and negative samples. We separate and disambiguate the same ad creative served on different days in the positive and negative samples. We found 46\% of the ambiguous samples, and we finally eliminated the ambiguous samples. So far we have completed the training data of the autoco model, but the model trained with such data will produce obvious positive and negative sample deviation phenomenon. Therefore, we cascade the list-wise distillent, the distilled rank model samples all samples for training, and takes whether the click is gt, and the rerank model uses the rank model to distill to obtain the label.

\section{Method}\label{sec3}
The platform framework of the dynamic creative optimization algorithm proposed in this paper is fig7 as follows:

\begin{figure}[h]%
\centering
\includegraphics[width=0.9\textwidth]{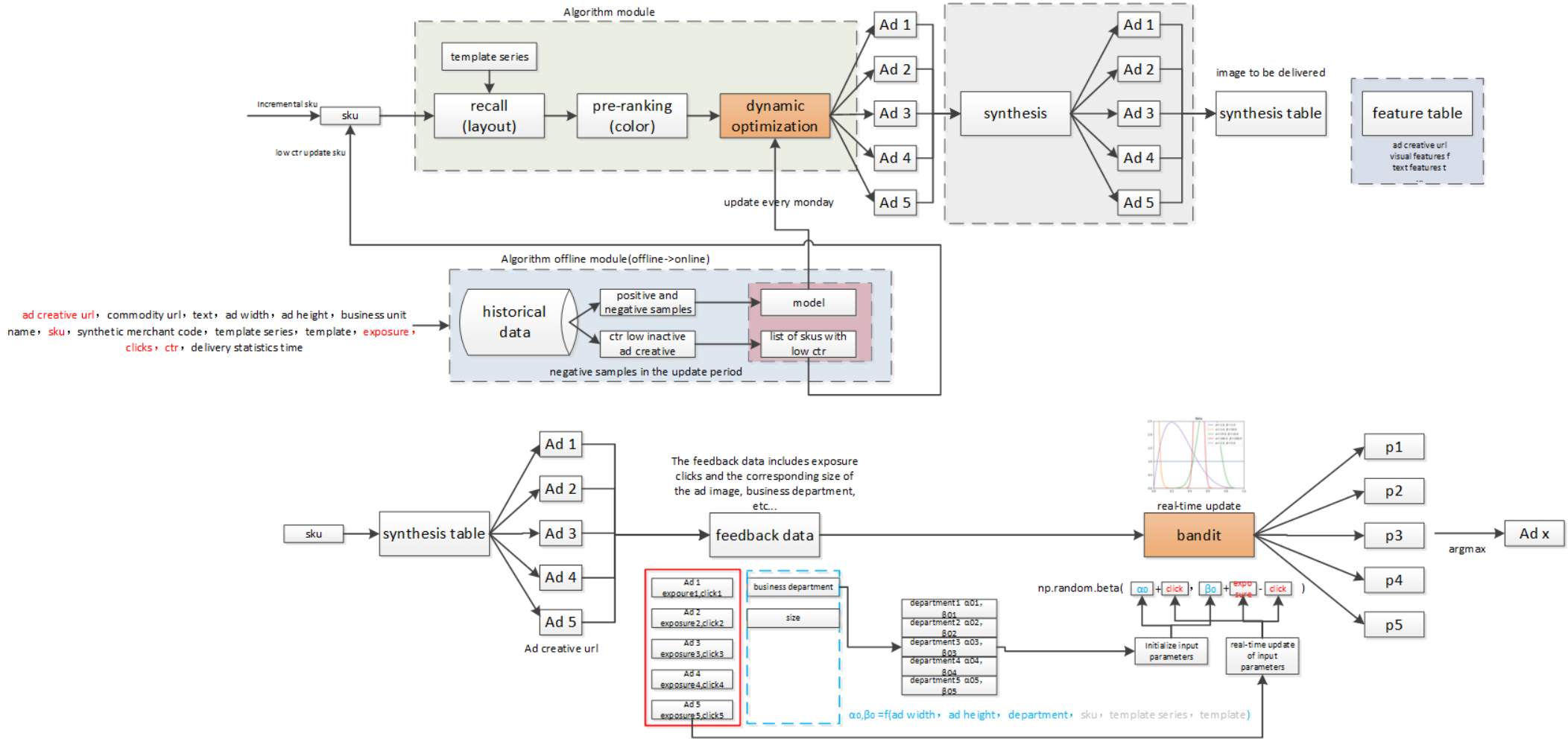}
\caption{\centering the platform framework of the dynamic creative optimization algorithm}\label{fig7}
\end{figure}

The first stage of our method is embedded in creative optimization as a whole, the second stage is a bandit, and the whole is a two-stage dynamic creative optimization framework.As shown in fig8.

\begin{figure}[h]%
\centering
\includegraphics[width=0.9\textwidth]{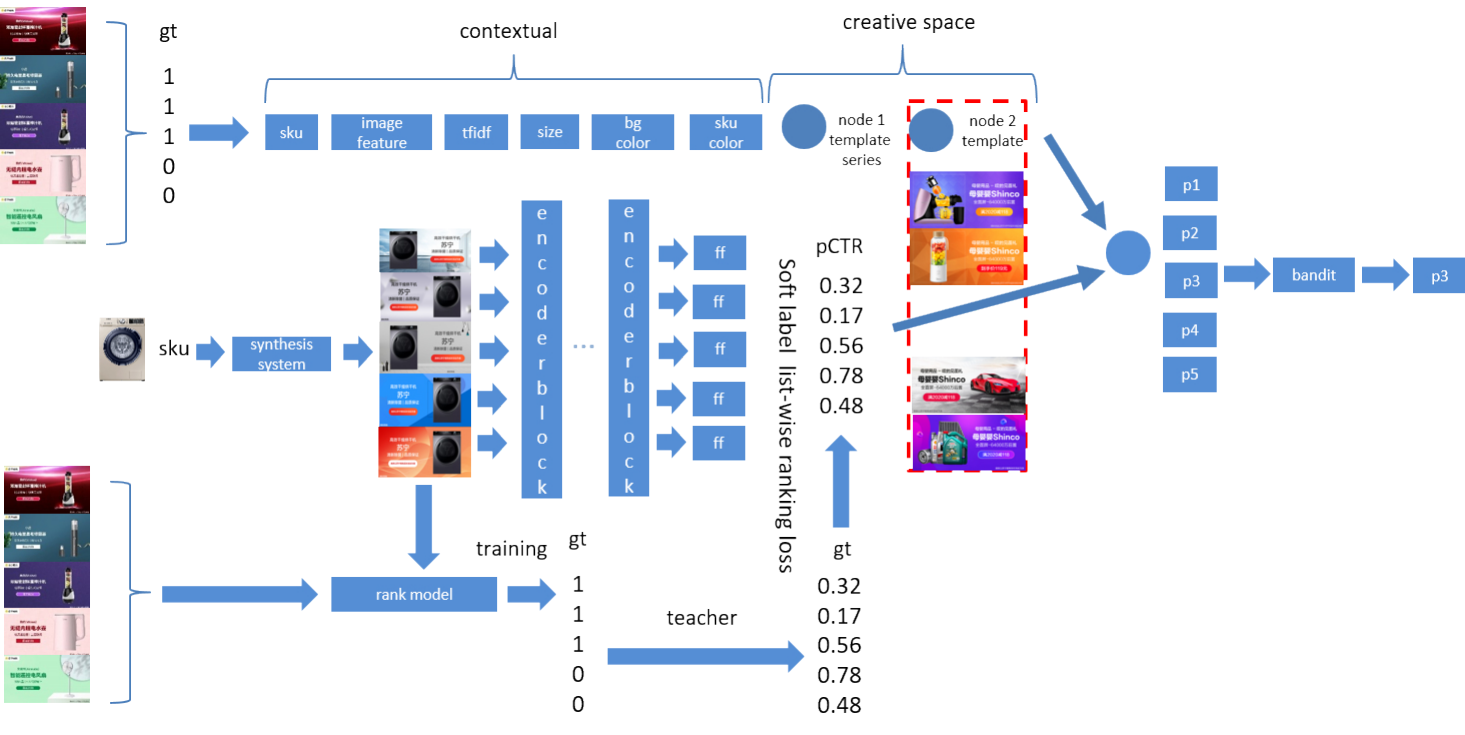}
\caption{\centering the dynamic creative selectio}\label{fig8}
\end{figure}

In our approach we decouple this dynamic creative optimization into two stages, a cascaded structure that can trade off between effectiveness and efficiency. In the first stage, we train an automatic creative optimization architecture based on autoco to simulate complex interactions between creative elements. After autoco, although we obtained the sorting of different creative images under a sku. However, due to the removal of ambiguous samples and sparse samples, the model itself is not unbiased. We propose a transformer-based rerank model. With the help of the rank model, we propose a distillation method to learn the relative order of ideas and extract the ranking knowledge to guide rerank learning. The creative sequence soft labels in each advertisement are generated by a rank model to alleviate the dilemma that a large number of under-displayed creatives cannot obtain real labels. In the second stage, we design a bandit model, and bandit selects one of the five images for timely delivery. We propose a two-stage framework, one-stage modeling between child nodes, strengthening feature crossover, screening better feature combinations, two-stage bandit, highlighting the solution to the E\&E problem. The first stage is n choose 5, the second stage is 5 choose 1.

\subsection{the first stage}\label{subsec4}

\subsubsection{autoco model}\label{subsec4}
The Autoco model borrows and modifies the method proposed by alibaba, and our element collection includes merchandise, text, size, template series, and templates. Usually FM is used, using the inner product as the interaction function and succeeds in ctr prediction, however due to the complexity of the interaction, the inner product may not achieve the best performance. To encourage complex interactions between different features, autoco chooses a combination of concat, multiply, plus, max and min. Similar to sif\cite{bib16}, autoco develops a fully connected layer for each function, and fc controls the output size of different operations to be consistent. autoco found that the selected 5 operations all had better performance than the fm model, with different degrees of optimization. There are more approximate interaction functions between different elements. After the interaction function is completed, it will be transformed into a latent space of the same size through a layer of full connection. In order to obtain the optimal interaction function between different features, a straightforward idea is to traverse all possibilities, but the time complexity in this case is exponential. In order to search for the interaction function between features more quickly, we use a one-shot search algorithm borrowed from the algorithm in NASP \cite{bib17}. Feature dimensions are characterized before entering autoco.As shown in Fig9.

\begin{figure}[h]%
\centering
\includegraphics[width=0.9\textwidth]{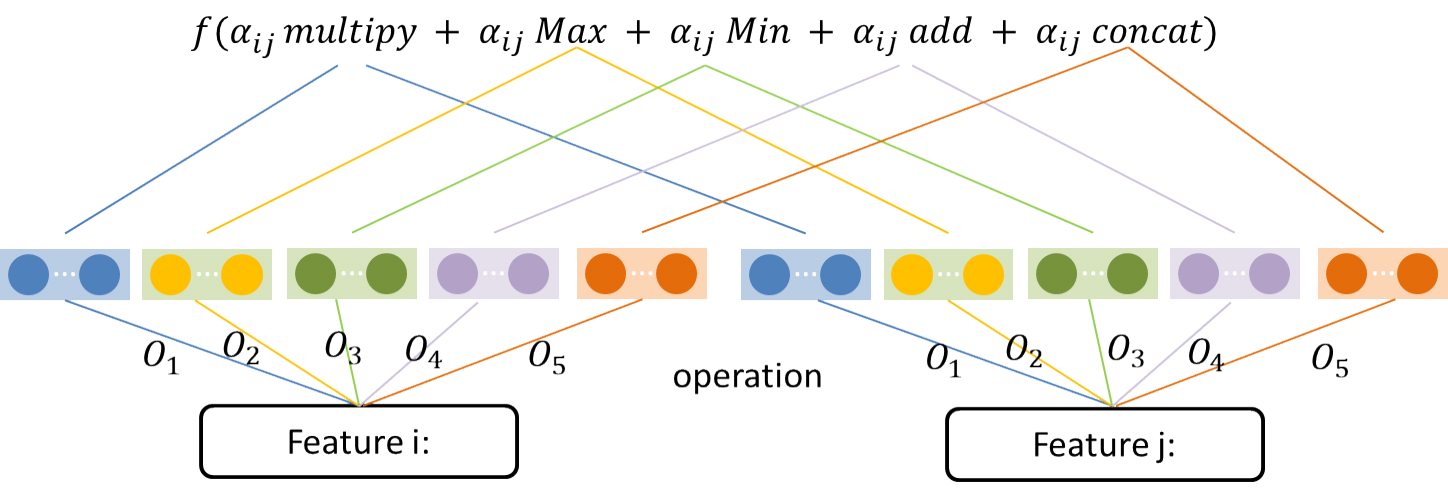}
\caption{\centering interactive function search space representation}\label{fig9}
\end{figure}

We divided into two stages in the feature dimension of autoco, the first dimension is contextual, and the second dimension is creative space. In contextual, we selected sku, image feature, tfidf, size, bg color and sku color, and selected template series and template in the creative space. Template series and template are the dimensions of creative screening.

a. Image feature extraction

Both autoencoder self-supervised and supervised feature extraction are used. Autoencoder adopts the encoder-decoder structure, and the supervised training adopts the division as the supervision label. We used advertising creative of 27 divisions, with a total of 312,925, including 250,340 training data and 62,585 test data.

\begin{figure}[h]%
\centering
\includegraphics[width=0.9\textwidth]{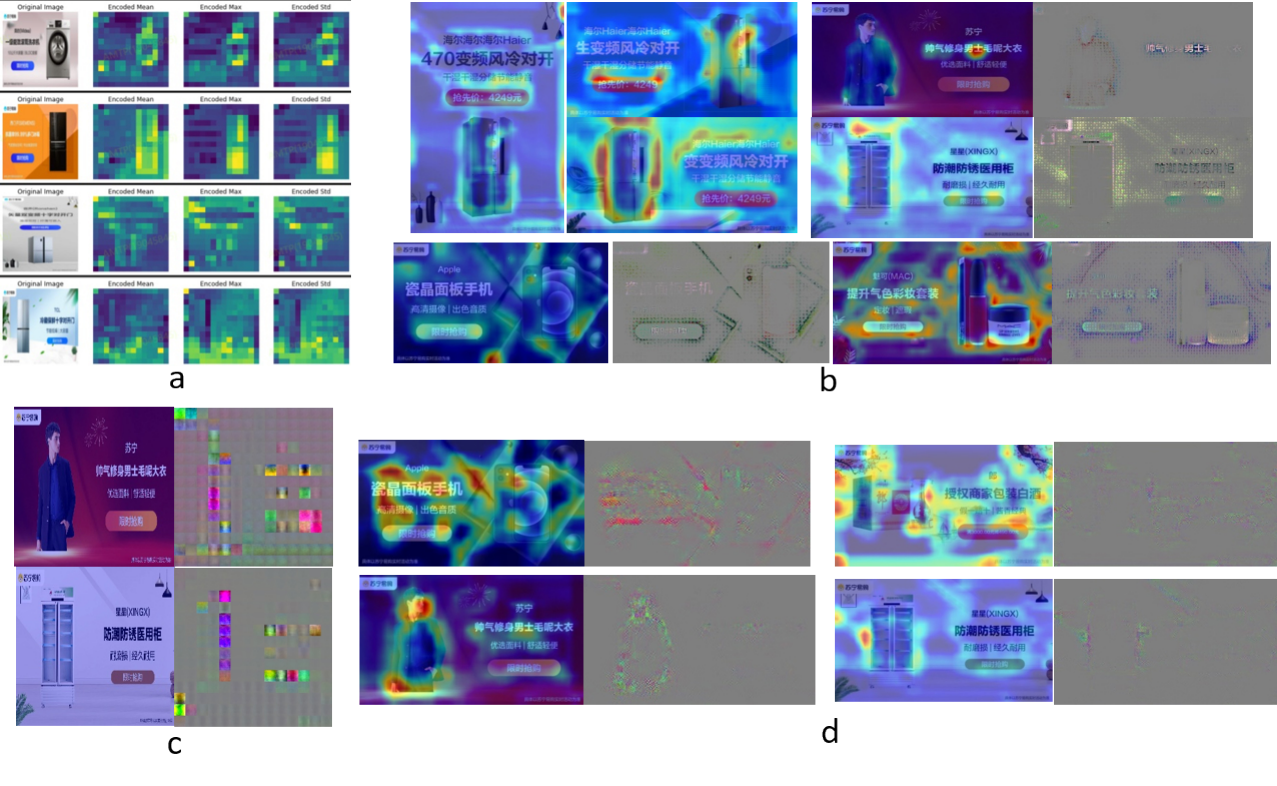}
\caption{\centering cam image of feature extraction}\label{fig10}
\end{figure}

As shown in Fig10.a is the cam image of autoencoder unsupervised feature extraction. b is the cam image of the supervised feature extraction of res2net50. It can be clearly seen that the feature extraction of res2net is relatively fine. c is the cam image of vit's supervised feature extraction, vit is a fixed-size input, and his feature extraction is relatively random, which is also discussed a lot. d is the feature extraction cam map of mbilenetv2, we tried the extraction of lightweight network, and finally we chose the feature extraction method of res2net.

b. Text dimension feature extraction

The text dimension uses tfidf as word frequency feature embedding, and we also tried finetune bert for downstream text feature extraction. However, since the mode of the text is relatively fixed and there is no long continuous semantic information, the bert effect of finetune is general. In the end, we only chose tfidf to extract word frequency. In addition, we did one-hot encoding for some words such as "discount" and "full discount" in the text.

c. Id class feature modeling

The main color is extracted, and 12 main colors are artificially designed, including black, light, red, orange, yellow, yellow-green, green, turquoise, cyan, indigo, blue, purple, magenta, and purple. Numerical feature dimensions such as size and id are considered. For example, seriesId is the template series, id is the template number, and department is the business department number of the product group.

The training metrics for Autoco in Table3.We finally chose fm\_nas\_fc, which has higher recall on positive samples.

\begin{table}[h]
\begin{center}
\begin{minipage}{\textwidth}
\caption{\centering the training metrics for Autoco }\label{tab1}%
\resizebox{\textwidth}{70mm}{

\begin{tabular}{@{}lccccccc@{}}
\midrule
model& logloss & acc & auc & & precision & recall & f1-score\\
\midrule

\multirow{2}{*}{fm} & \multirow{2}{*}{0.5856} & \multirow{2}{*}{0.646} & \multirow{2}{*}{0.6589} & 0 & 0.67 & 0.81 & 0.73 \\
\cline{5-8}
& & & &  1 & 0.58 & 0.41 & 0.48 \\
\midrule

\multirow{2}{*}{fm\_ts} & \multirow{2}{*}{0.6281} & \multirow{2}{*}{0.649} & \multirow{2}{*}{0.672} & 0 & 0.66 & 0.84 & 0.74 \\
\cline{5-8}
& & & &  1 & 0.60 & 0.37 & 0.46 \\
\midrule

\multirow{2}{*}{fm\_nas} & \multirow{2}{*}{0.677} & \multirow{2}{*}{0.633} & \multirow{2}{*}{0.637} & 0 & 0.67 & 0.78 & 0.72 \\
\cline{5-8}
& & & &  1 & 0.56 & 0.41 & 0.48 \\
\midrule

\multirow{2}{*}{fm\_nas\_fc} & \multirow{2}{*}{0.6487} & \multirow{2}{*}{0.649} & \multirow{2}{*}{0.657} & 0 & 0.67 & 0.80 & 0.73 \\
\cline{5-8}
& & & &  1 & 0.59 & 0.42 & 0.49 \\
\midrule

\multirow{2}{*}{fm\_nas\_oae} & \multirow{2}{*}{0.6446} & \multirow{2}{*}{0.644} & \multirow{2}{*}{0.658} & 0 & 0.67 & 0.81 & 0.73 \\
\cline{5-8}
& & & &  1 & 0.58 & 0.40 & 0.47 \\
\midrule

\multirow{2}{*}{fm\_nas\_oae\_fc} & \multirow{2}{*}{0.6507} & \multirow{2}{*}{0.644} & \multirow{2}{*}{0.9,656} & 0 & 0.67 & 0.80 & 0.73 \\
\cline{5-8}
& & & &  1 & 0.58 & 0.41 & 0.48 \\
\midrule

\multirow{2}{*}{fm\_snas\_fc} & \multirow{2}{*}{1.283} & \multirow{2}{*}{0.593} & \multirow{2}{*}{0.553} & 0 & 0.61 & 0.81 & 0.73 \\
\cline{5-8}
& & & &  1 & 0.59 & 0.39 & 0.47 \\
\midrule

\multirow{2}{*}{fm\_dsnas\_fc} & \multirow{2}{*}{0.6414} & \multirow{2}{*}{0.645} & \multirow{2}{*}{0.658} & 0 & 0.67 & 0.81 & 0.73 \\
\cline{5-8}
& & & &  1 & 0.59 & 0.39 & 0.47 \\
\midrule

\multirow{2}{*}{fm\_nas\_ts\_fc} & \multirow{2}{*}{0.6359} & \multirow{2}{*}{0.6532} & \multirow{2}{*}{0.665} & 0 & 0.67 & 0.82 & 0.74 \\
\cline{5-8}
& & & &  1 & 0.60 & 0.40 & 0.48 \\
\midrule

\multirow{2}{*}{fm\_snas\_ts\_fc} & \multirow{2}{*}{0.982} & \multirow{2}{*}{0.401} & \multirow{2}{*}{0.559} & 0 & 0.60 & 0.00 & 0.00 \\
\cline{5-8}
& & & &  1 & 0.40 & 1.00 & 0.57 \\
\midrule

\multirow{2}{*}{fm\_dsnas\_ts\_fc} & \multirow{2}{*}{0.637} & \multirow{2}{*}{0.63} & \multirow{2}{*}{0.661} & 0 & 0.66 & 0.82 & 0.73 \\
\cline{5-8}
& & & &  1 & 0.58 & 0.38 & 0.46 \\
\midrule

\end{tabular}}
\end{minipage}
\end{center}
\end{table}

\subsubsection{list-wise ranking distillation}\label{subsec4}
    
The creative fine-arrangement model needs to consider multi-modal dimension features, but since the input data does not have user-side features, the material dimension features may be repeated in different days. Strictly positive and negative samples are taken out by periodic bucketing, but there are several problems:1. This strictly positive and negative sample is actually designed to deviate from the original task, and duplicate sample data between the two cannot achieve a good result in the model.2. A large percentage of creatives cannot be fully displayed, in order to narrow the number of ads considered by bandit. The number of ad creatives is small, and bandit can easily achieve its optimal selection, so the one-stage recommendation should fully consider those sparse sample data.3. Creative selection only needs to learn the relative order of the creatives, not the ctr scores. Therefore, we treat this task as a learning ranking problem to rank the ad creative produced by the input sku. Inspired by ranking distillation, we propose a soft-label list ranking distillation method to extract ranking knowledge from teacher ranking model to guide student rerank model learning. Therefore, a rank model is trained with non-overlapping data, and the ctr predicted by the rank model is used as the dominant, and the order value of different synthetic images under a sku is obtained in the synthetic image frame. Train a learn2rank model that provides ranking relationships of different ideas for the same sku. Since the rank model fully considers sparse and ambiguous samples, it can well fuse the sample information, use the rank model to predict the click-through rate of each creative, and use the click-through rate score as a soft label for the order of creatives in the offline training phase. In essence, we associate the data in the fuzzy zone with the clear positive and negative samples through some kind of diffusion of learn2rank, and perform feature interaction.

a.	learn2rank model

The architecture of the learn2rank model can be thought of as the encoder part of the Transformer. After passing the input list through multiple encoder blocks, a shared fully connected layer assigns a final score to each item. Our learn2rank model is trained using 30,000 sku-distilled label sets, including 7.5w for training data and 73,270 images for test data.

The key building block of our model is the self-attention operation. Thanks to its formulation, all items present in the list are taken into account while computing the score for a given item, both in training and inference.

Scaled dot-product attention is defined as follows:

Suppose Q is a d-dimensional matrix representing all items in the list. Let K and V be the keys and values matrices, respectively.

\begin{center}
   $Attention(Q,K,V)=softmax(\frac{QK^{T}}{\sqrt{d} }  )V$ 
\end{center}

When Q=K=V, we call this operation self-attention.
\begin{figure}[h]%
\centering
\includegraphics[width=0.9\textwidth]{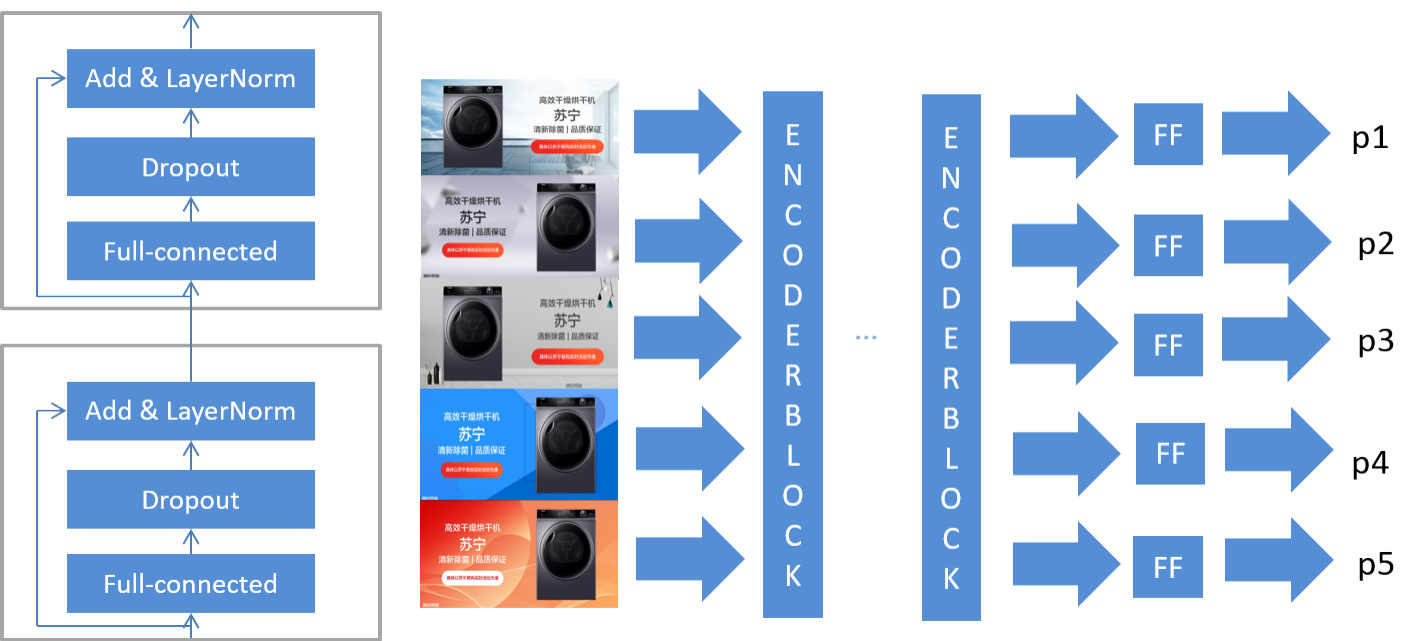}
\caption{\centering learn2rank model}\label{fig11}
\end{figure}

The model of learn2rank used in this paper is shown in Fig11,The training metrics of the learn2rank model in Table4.We finally chose the transformer-based lean2rank model.

\begin{table}[h]
\begin{center}
\begin{minipage}{\textwidth}
\caption{\centering the training metrics for learn2rank }\label{tab1}%
\resizebox{\textwidth}{16mm}{
\begin{tabular}{@{}lcccccccccccccc@{}}

\midrule
\multirow{2}{*}{fm}  &  NDCG@5 &  & \multirow{2}{*}{val loss} & \multirow{2}{*}{loss func} & \multirow{2}{*}{epoch} & \multirow{2}{*}{batch} & \multirow{2}{*}{lr} & \multirow{2}{*}{optimizer} & \multirow{2}{*}{with pe} & \multirow{2}{*}{N layers} & \multirow{2}{*}{H multi} & \multirow{2}{*}{diff} & \multirow{2}{*}{dropout} & \multirow{2}{*}{l} \\
\cline{2-3}
& best & current  \\
\midrule

\multirow{2}{*}{transformer} & 0.7204 & 0.7199 & no & ordinal & 188 & 960 & 0.001 & adam & null & 4 & 2 & 512 & 0.4 & 5 \\
\cline{2-14}
& 0.7191 & 0.7179 & no & ordinal & 101 & 960 & 0.0001 & adam & null & 4 & 2 & 512 & 0.4 & 5 \\
\midrule

\multirow{2}{*}{mlp} & 0.7190 & 0.7176 & yes & ordinal & 104 & 960 & 0.0001 & adam & null & 4 & 2 & 512 & 0.4 & 5 \\
\cline{2-14}
& 0.7191 & 0.7180 & yes & ordinal & 104 & 960 & 0.00001 & adam & null & 4 & 2 & 512 & 0.4 & 5 \\
\midrule

\end{tabular}}
\end{minipage}
\end{center}
\end{table}

b.	rank model

The Rank model mainly considers multiple ctr prediction models. We refer to the deepctr\_torch library for train.The results are shown in Table5,we chose the autoint model in the end .

\begin{table}[h]
\begin{center}
\begin{minipage}{\textwidth}
\caption{\centering the training metrics for ctr estimate }\label{tab1}%
\resizebox{\textwidth}{50mm}{

\begin{tabular}{@{}lccccccc@{}}
\midrule
model& logloss & acc & auc & & precision & recall & f1-score\\
\midrule

\multirow{2}{*}{deepfm} & \multirow{2}{*}{0.9239} & \multirow{2}{*}{0.601} & \multirow{2}{*}{0.625} & 0 & 0.63 & 0.67 & 0.65 \\
\cline{5-8}
& & & &  1 & 0.57 & 0.52 & 0.55 \\
\midrule

\multirow{2}{*}{wdl} & \multirow{2}{*}{0.8491} & \multirow{2}{*}{0.0.599} & \multirow{2}{*}{0.6274} & 0 & 0.63 & 0.64 & 0.63 \\
\cline{5-8}
& & & &  1 & 0.56 & 0.55 & 0.55 \\
\midrule

\multirow{2}{*}{dcn} & \multirow{2}{*}{0.762} & \multirow{2}{*}{0.609} & \multirow{2}{*}{0.6333} & 0 & 0.62 & 0.74 & 0.67 \\
\cline{5-8}
& & & &  1 & 0.59 & 0.46 & 0.52 \\
\midrule

\multirow{2}{*}{xDeepFM} & \multirow{2}{*}{0.9628} & \multirow{2}{*}{0.604} & \multirow{2}{*}{0.6285} & 0 & 0.63 & 0.65 & 0.64 \\
\cline{5-8}
& & & &  1 & 0.57 & 0.55 & 0.56 \\
\midrule

\multirow{2}{*}{nfm} & \multirow{2}{*}{0.7189} & \multirow{2}{*}{0.614} & \multirow{2}{*}{0.6385} & 0 & 0.62 & 0.75 & 0.68 \\
\cline{5-8}
& & & &  1 & 0.69 & 0.45 & 0.52 \\
\midrule

\multirow{2}{*}{pnn} & \multirow{2}{*}{0.7763} & \multirow{2}{*}{0.61} & \multirow{2}{*}{0.6273} & 0 & 0.62 & 0.71 & 0.66 \\
\cline{5-8}
& & & &  1 & 0.59 & 0.49 & 0.53 \\
\midrule

\multirow{2}{*}{fibinet} & \multirow{2}{*}{0.7869} & \multirow{2}{*}{0.6132} & \multirow{2}{*}{0.634} & 0 & 0.62 & 0.74 & 0.68 \\
\cline{5-8}
& & & &  1 & 0.60 & 0.46 & 0.52 \\
\midrule

\multirow{2}{*}{afm} & \multirow{2}{*}{0.667} & \multirow{2}{*}{0.623} & \multirow{2}{*}{0.6527} & 0 & 0.64 & 0.71 & 0.67 \\
\cline{5-8}
& & & &  1 & 0.60 & 0.52 & 0.56 \\
\midrule

\multirow{2}{*}{autoint} & \multirow{2}{*}{0.8598} & \multirow{2}{*}{0.602} & \multirow{2}{*}{0.6313} & 0 & 0.64 & 0.61 & 0.63 \\
\cline{5-8}
& & & &  1 & 0.56 & 0.59 & 0.57 \\
\midrule

\end{tabular}}
\end{minipage}
\end{center}
\end{table}

c. distillation

In an ad, there are multiple creative scores ${s_{1},s_{2},...s_{i},...,s_{m} }$ and their corresponding soft labels ${y_{1},y_{2},...y_{i},...,y_{m}}$. We directly use soft labels to supervise the learn2rank model.

\subsection{the second stage}\label{subsec4}
Epsilon-greedy\cite{bib2}, Thompson sampling\cite{bib3} and UCB\cite{bib4} are classic context-free algorithms. They use the reward/cost from the model to update their E\&E policy without contextual information. Because web content changes frequently, it is difficult for models to quickly adapt to new ideas. \cite{bib18} extend these context-free methods by considering auxiliary information such as user/content representations. They assume that the expected return of the arm is linear in its characteristics. The main problem facing linear algorithms is that they lack representational power, they complement accurate uncertainty estimates. An impression occurs when an ad creative is shown to a user, and $R_{C_{t} } =1$ means the user clicked on the ad creative. The goal of a multi\-armed bandit is to minimize accumulated regret within T steps:

\begin{center}
     $min\sum_{T}^{t}(R_{C_{*} } -R_{C_{t} } )$ 
\end{center}

where $c=argmax(E(R_{c} ))$ represents the candidate idea with the largest expected return. We ended up choosing thompson.

\section{Experiments}\label{sec5}
 
\subsection{metrics}\label{subsec5}
In offline experiments, we use simulated click-through rate (sctr)\cite{bib11} as an evaluation metric. sctr is a metric used to simulate the online performance of creative selection. For an impression, we forecast all the creatives under the ad and select the best creative based on the forecast score. If the selected creative is the same as the display creative for this record, we consider this a valid impression. impression = impression+1,clicks=clicks+y,where y indicates the label of the actual click, and the sctr is defined as sctr=clicks/impression.

\subsection{evaluation system}\label{subsec5}
In our creative optimization framework, we have customized a complete set of evaluation systems, including offline evaluation of the first and second stages. In the online stage, the ctr of the whole process is evaluated separately. The complete process includes the creative generation system of recall, pre-ranking and creative ranking, and the advertising creative push system including bandit. In the creative ranking, we use the creative generation model jointly developed by autoco and learn2rank. The generated creative is evaluated by generating ctr through the rank model on the offline side. The rank model is used as a unified model predictor. Here we use autoint as the ctr prediction model.Evaluation system as shown in Fig12.

\begin{figure}[h]%
\centering
\includegraphics[width=0.9\textwidth]{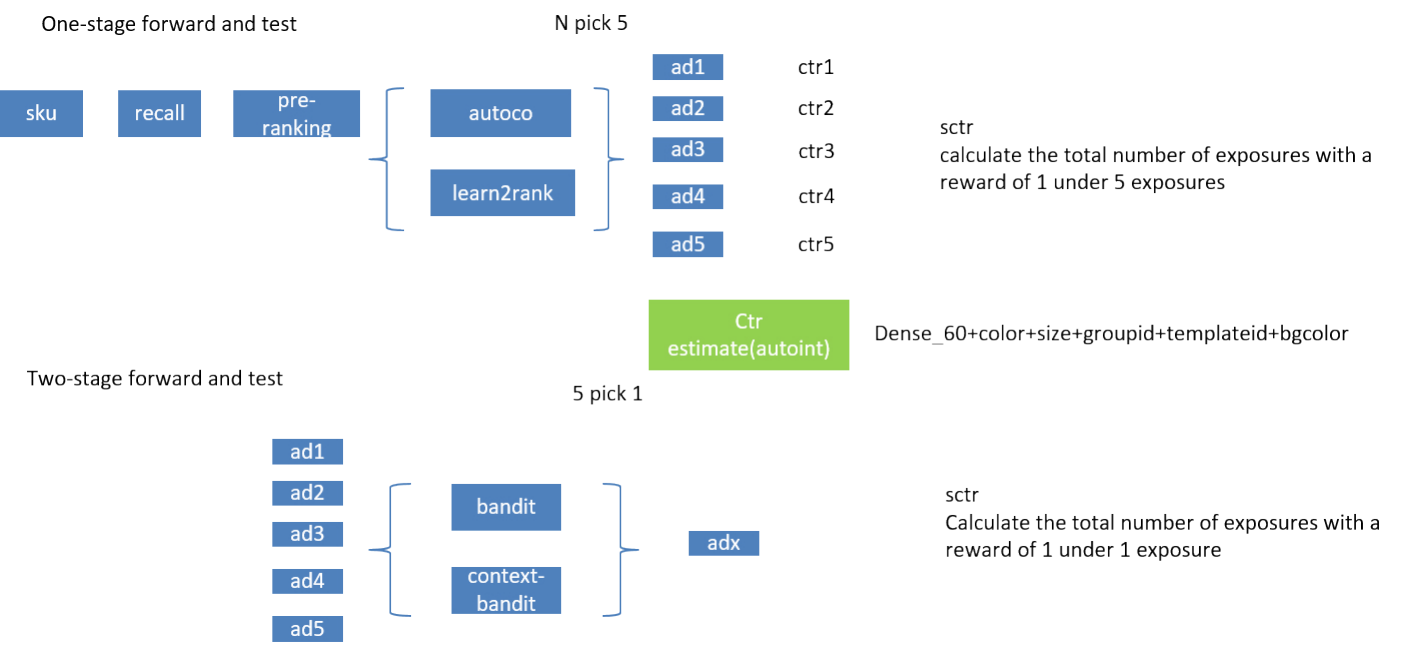}
\caption{\centering creative optimization frameworkl}\label{fig12}
\end{figure}

\subsection{model training metrics}\label{subsec5}
\subsubsection{the first stage evaluation metrics}
\begin{figure}[h]%
\centering
\includegraphics[width=0.9\textwidth]{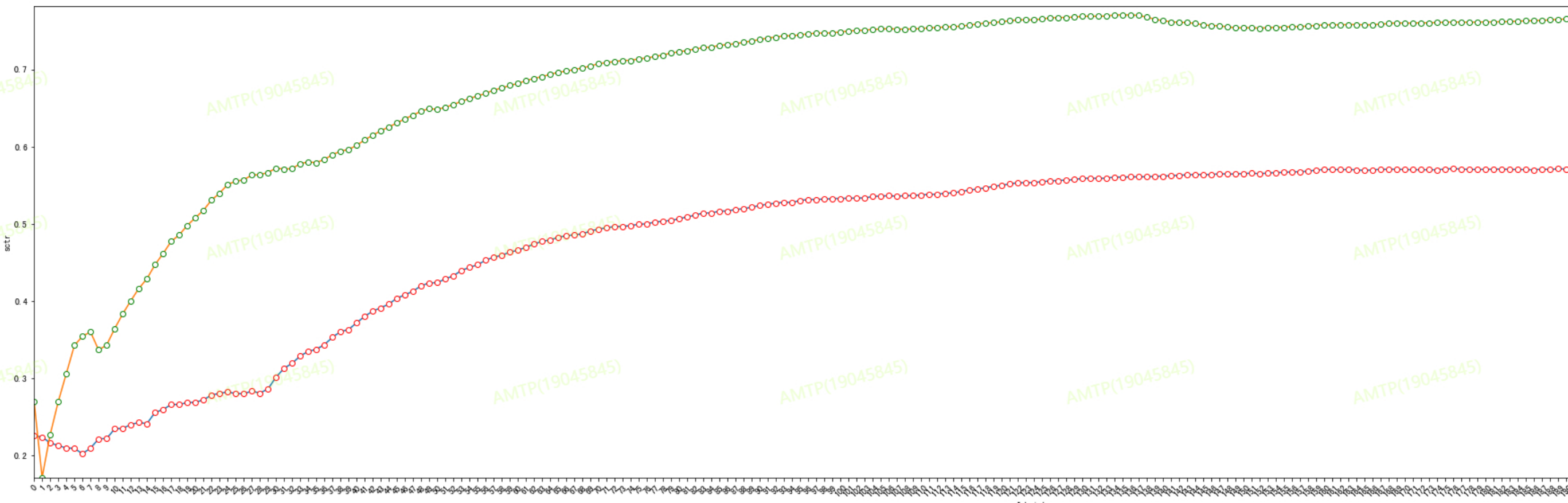}
\caption{\centering the first stage evaluation curve}\label{fig13}
\end{figure}

In the first stage, we tested both lgb and autoco methods. It can be seen that the effect of autoco is obviously better. The green line represents autoco, and the red line represents lgb. The horizontal axis is the number of exposures, and the vertical axis is the sctr.

\subsubsection{the second stage evaluation metrics}
\begin{figure}[h]%
\centering
\includegraphics[width=0.9\textwidth]{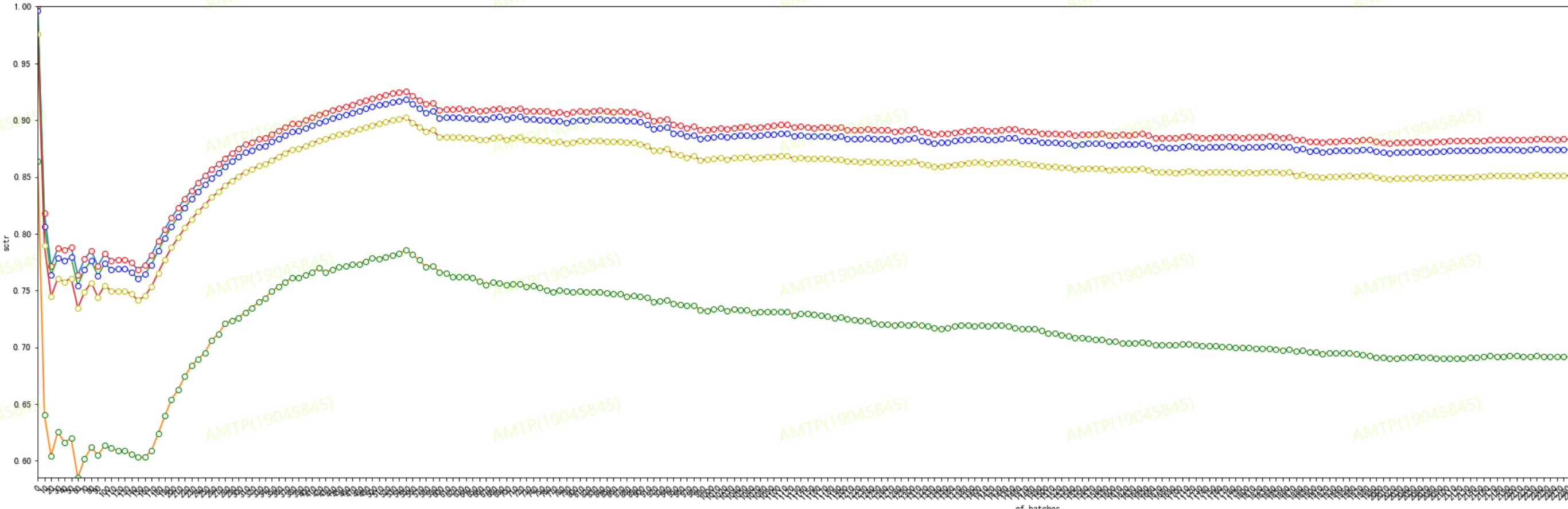}
\caption{\centering the second stage evaluation curve}\label{fig14}
\end{figure}

The red line is autoco-opt, that is, in the forward reasoning process, n chooses 5, and it is the image with the highest score in the rank model. It can be seen that sctr is the highest in this case. The blue line is autoco-ts, the yellow line is autoco-hbm, and the green line is auto-random.

\subsection{online result display}\label{subsec5}
\begin{figure}[h]%
\centering
\includegraphics[width=0.9\textwidth]{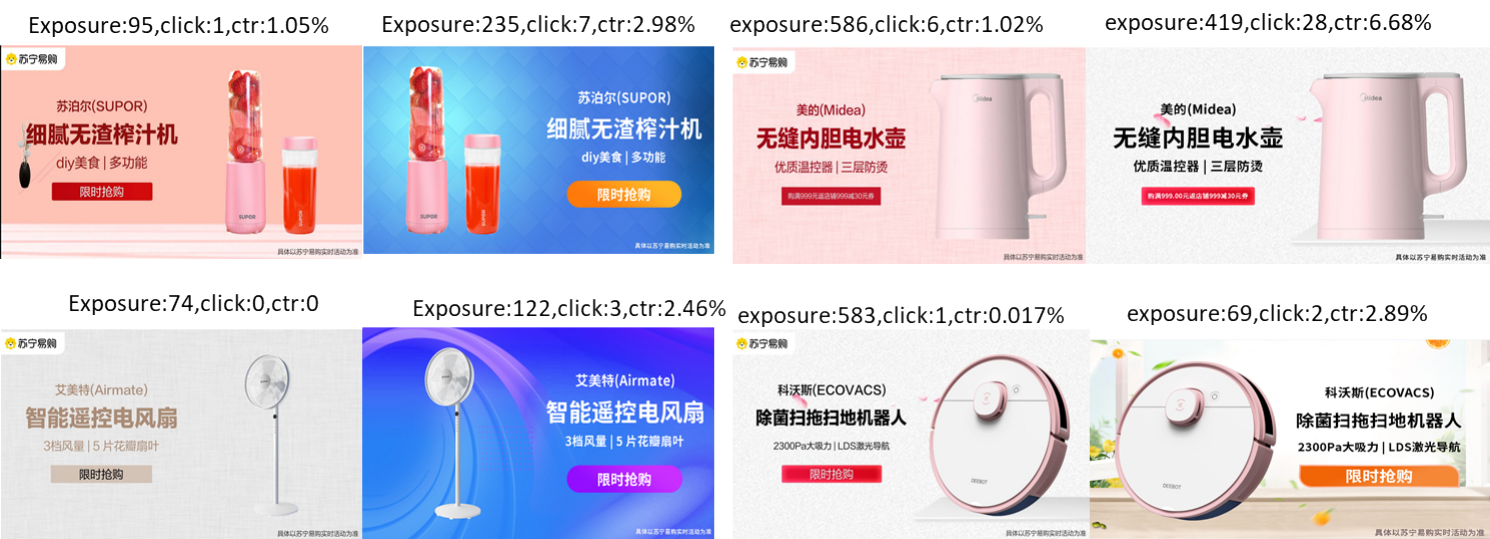}
\caption{\centering ad creative comparison after intelligent optimization online}\label{fig15}
\end{figure}

Obviously, the ctr of the image after the creative optimization has become higher, and the contrast and matching have been significantly improved.

\section{Conclusion}\label{sec5}
In our approach we decouple this dynamic creative optimization into two stages. In the first stage, we train an automatic creative optimization architecture based on autoco to simulate complex interactions between creative elements. In the first stage, we train an automatic creative optimization architecture based on autoco to simulate complex interactions between creative elements. Extend the multiplication operator in the inner product to the operator set {concat,multiply,plus,max,min} through the idea of automl. After autoco, we obtained the ranking of different creative under a sku. Considering the sample ambiguity and sparseness, we proposed a transformer-based learn2rank model. With the help of the rank model, a distillation method is proposed to learn the relative order of creatives and extract the ranking knowledge to guide learn2rank learning. The creative sequence soft labels in each advertisement are generated by the rank model to alleviate the dilemma that a large number of under-displayed creatives cannot obtain real labels. In the second stage, we chose the bandit model in a smaller creative space. Based on the first stage, we output five images based on the same sku. In the second stage, we designed a bandit model, and the bandit selected one of the five images for timely delivery. It has made significant progress in suning's out-of-feed scenario.

\section*{Compliance with Ethical Standards}
This article does not contain any studies with human participants or animals performed by any of the authors.In this experiment, we did not collect any samples of human and animals.

\section*{Competing Interests}
The authors declared that they have no conflicts of interest to this work.
We declare that we do not have any commercial or associative interest that represents a conflict of interest in connection with the work submitted.

\section*{Research Data Policy and Data Availability Statements}
Because design resources involve company assets and online creative preference data involves privacy concerns, participants in this study did not consent to the public sharing of their data, so supporting data was not available.

%%\bibliography{sn-bibliography}% common bib file

\begin{thebibliography}{18}
\ifx \bisbn   \undefined \def \bisbn  #1{ISBN #1}\fi
\ifx \binits  \undefined \def \binits#1{#1}\fi
\ifx \bauthor  \undefined \def \bauthor#1{#1}\fi
\ifx \batitle  \undefined \def \batitle#1{#1}\fi
\ifx \bjtitle  \undefined \def \bjtitle#1{#1}\fi
\ifx \bvolume  \undefined \def \bvolume#1{\textbf{#1}}\fi
\ifx \byear  \undefined \def \byear#1{#1}\fi
\ifx \bissue  \undefined \def \bissue#1{#1}\fi
\ifx \bfpage  \undefined \def \bfpage#1{#1}\fi
\ifx \blpage  \undefined \def \blpage #1{#1}\fi
\ifx \burl  \undefined \def \burl#1{\textsf{#1}}\fi
\ifx \doiurl  \undefined \def \doiurl#1{\url{https://doi.org/#1}}\fi
\ifx \betal  \undefined \def \betal{\textit{et al.}}\fi
\ifx \binstitute  \undefined \def \binstitute#1{#1}\fi
\ifx \binstitutionaled  \undefined \def \binstitutionaled#1{#1}\fi
\ifx \bctitle  \undefined \def \bctitle#1{#1}\fi
\ifx \beditor  \undefined \def \beditor#1{#1}\fi
\ifx \bpublisher  \undefined \def \bpublisher#1{#1}\fi
\ifx \bbtitle  \undefined \def \bbtitle#1{#1}\fi
\ifx \bedition  \undefined \def \bedition#1{#1}\fi
\ifx \bseriesno  \undefined \def \bseriesno#1{#1}\fi
\ifx \blocation  \undefined \def \blocation#1{#1}\fi
\ifx \bsertitle  \undefined \def \bsertitle#1{#1}\fi
\ifx \bsnm \undefined \def \bsnm#1{#1}\fi
\ifx \bsuffix \undefined \def \bsuffix#1{#1}\fi
\ifx \bparticle \undefined \def \bparticle#1{#1}\fi
\ifx \barticle \undefined \def \barticle#1{#1}\fi
\bibcommenthead
\ifx \bconfdate \undefined \def \bconfdate #1{#1}\fi
\ifx \botherref \undefined \def \botherref #1{#1}\fi
\ifx \url \undefined \def \url#1{\textsf{#1}}\fi
\ifx \bchapter \undefined \def \bchapter#1{#1}\fi
\ifx \bbook \undefined \def \bbook#1{#1}\fi
\ifx \bcomment \undefined \def \bcomment#1{#1}\fi
\ifx \oauthor \undefined \def \oauthor#1{#1}\fi
\ifx \citeauthoryear \undefined \def \citeauthoryear#1{#1}\fi
\ifx \endbibitem  \undefined \def \endbibitem {}\fi
\ifx \bconflocation  \undefined \def \bconflocation#1{#1}\fi
\ifx \arxivurl  \undefined \def \arxivurl#1{\textsf{#1}}\fi
\csname PreBibitemsHook\endcsname

%%% 1
\bibitem{bib1}
\begin{bchapter}
\bauthor{\bsnm{Chen}, \binits{J.}},
\bauthor{\bsnm{Sun}, \binits{B.}},
\bauthor{\bsnm{Li}, \binits{H.}},
\bauthor{\bsnm{Lu}, \binits{H.}},
\bauthor{\bsnm{Hua}, \binits{X.-S.}}:
\bctitle{Deep ctr prediction in display advertising}.
In: \bbtitle{Proceedings of the 24th ACM International Conference on
  Multimedia},
pp. \bfpage{811}--\blpage{820}
(\byear{2016})
\end{bchapter}
\endbibitem

%%% 2
\bibitem{bib2}
\begin{barticle}
\bauthor{\bsnm{Fran{\c{c}}ois-Lavet}, \binits{V.}},
\bauthor{\bsnm{Henderson}, \binits{P.}},
\bauthor{\bsnm{Islam}, \binits{R.}},
\bauthor{\bsnm{Bellemare}, \binits{M.G.}},
\bauthor{\bsnm{Pineau}, \binits{J.}}, \betal:
\batitle{An introduction to deep reinforcement learning}.
\bjtitle{Foundations and Trends{\textregistered} in Machine Learning}
\bvolume{11}(\bissue{3-4}),
\bfpage{219}--\blpage{354}
(\byear{2018})
\end{barticle}
\endbibitem

%%% 3
\bibitem{bib3}
\begin{barticle}
\bauthor{\bsnm{Russo}, \binits{D.}},
\bauthor{\bsnm{Van~Roy}, \binits{B.}}:
\batitle{Learning to optimize via posterior sampling}.
\bjtitle{Mathematics of Operations Research}
\bvolume{39}(\bissue{4}),
\bfpage{1221}--\blpage{1243}
(\byear{2014})
\end{barticle}
\endbibitem

%%% 4
\bibitem{bib4}
\begin{bchapter}
\bauthor{\bsnm{Bouneffouf}, \binits{D.}}:
\bctitle{Finite-time analysis of the multi-armed bandit problem with known
  trend}.
In: \bbtitle{2016 IEEE Congress on Evolutionary Computation (CEC)},
pp. \bfpage{2543}--\blpage{2549}
(\byear{2016}).
\bcomment{IEEE}
\end{bchapter}
\endbibitem

%%% 5
\bibitem{bib5}
\begin{bchapter}
\bauthor{\bsnm{Chen}, \binits{J.}},
\bauthor{\bsnm{Xu}, \binits{J.}},
\bauthor{\bsnm{Jiang}, \binits{G.}},
\bauthor{\bsnm{Ge}, \binits{T.}},
\bauthor{\bsnm{Zhang}, \binits{Z.}},
\bauthor{\bsnm{Lian}, \binits{D.}},
\bauthor{\bsnm{Zheng}, \binits{K.}}:
\bctitle{Automated creative optimization for e-commerce advertising}.
In: \bbtitle{Proceedings of the Web Conference 2021},
pp. \bfpage{2304}--\blpage{2313}
(\byear{2021})
\end{bchapter}
\endbibitem

%%% 6
\bibitem{bib6}
\begin{botherref}
\oauthor{\bsnm{Pobrotyn}, \binits{P.}},
\oauthor{\bsnm{Bartczak}, \binits{T.}},
\oauthor{\bsnm{Synowiec}, \binits{M.}},
\oauthor{\bsnm{Bia{\l}obrzeski}, \binits{R.}},
\oauthor{\bsnm{Bojar}, \binits{J.}}:
Context-aware learning to rank with self-attention.
arXiv preprint arXiv:2005.10084
(2020)
\end{botherref}
\endbibitem

%%% 7
\bibitem{bib7}
\begin{bchapter}
\bauthor{\bsnm{Lin}, \binits{K.}},
\bauthor{\bsnm{Zhang}, \binits{X.}},
\bauthor{\bsnm{Li}, \binits{F.}},
\bauthor{\bsnm{Wang}, \binits{P.}},
\bauthor{\bsnm{Long}, \binits{Q.}},
\bauthor{\bsnm{Deng}, \binits{H.}},
\bauthor{\bsnm{Xu}, \binits{J.}},
\bauthor{\bsnm{Zheng}, \binits{B.}}:
\bctitle{Joint optimization of ad ranking and creative selection}.
In: \bbtitle{Proceedings of the 45th International ACM SIGIR Conference on
  Research and Development in Information Retrieval},
pp. \bfpage{2341}--\blpage{2346}
(\byear{2022})
\end{bchapter}
\endbibitem

%%% 8
\bibitem{bib8}
\begin{bchapter}
\bauthor{\bsnm{Kang}, \binits{S.}},
\bauthor{\bsnm{Hwang}, \binits{J.}},
\bauthor{\bsnm{Kweon}, \binits{W.}},
\bauthor{\bsnm{Yu}, \binits{H.}}:
\bctitle{De-rrd: A knowledge distillation framework for recommender system}.
In: \bbtitle{Proceedings of the 29th ACM International Conference on
  Information \& Knowledge Management},
pp. \bfpage{605}--\blpage{614}
(\byear{2020})
\end{bchapter}
\endbibitem

%%% 9
\bibitem{bib9}
\begin{bchapter}
\bauthor{\bsnm{Zhao}, \binits{Z.}},
\bauthor{\bsnm{Li}, \binits{L.}},
\bauthor{\bsnm{Zhang}, \binits{B.}},
\bauthor{\bsnm{Wang}, \binits{M.}},
\bauthor{\bsnm{Jiang}, \binits{Y.}},
\bauthor{\bsnm{Xu}, \binits{L.}},
\bauthor{\bsnm{Wang}, \binits{F.}},
\bauthor{\bsnm{Ma}, \binits{W.}}:
\bctitle{What you look matters? offline evaluation of advertising creatives for
  cold-start problem}.
In: \bbtitle{Proceedings of the 28th ACM International Conference on
  Information and Knowledge Management},
pp. \bfpage{2605}--\blpage{2613}
(\byear{2019})
\end{bchapter}
\endbibitem

%%% 10
\bibitem{bib10}
\begin{bchapter}
\bauthor{\bsnm{Chen}, \binits{J.}},
\bauthor{\bsnm{Ge}, \binits{T.}},
\bauthor{\bsnm{Jiang}, \binits{G.}},
\bauthor{\bsnm{Zhang}, \binits{Z.}},
\bauthor{\bsnm{Lian}, \binits{D.}},
\bauthor{\bsnm{Zheng}, \binits{K.}}:
\bctitle{Efficient optimal selection for composited advertising creatives with
  tree structure}.
In: \bbtitle{Proceedings of the AAAI Conference on Artificial Intelligence},
vol. \bseriesno{35},
pp. \bfpage{3967}--\blpage{3975}
(\byear{2021})
\end{bchapter}
\endbibitem

%%% 11
\bibitem{bib11}
\begin{bchapter}
\bauthor{\bsnm{Wang}, \binits{S.}},
\bauthor{\bsnm{Liu}, \binits{Q.}},
\bauthor{\bsnm{Ge}, \binits{T.}},
\bauthor{\bsnm{Lian}, \binits{D.}},
\bauthor{\bsnm{Zhang}, \binits{Z.}}:
\bctitle{A hybrid bandit model with visual priors for creative ranking in
  display advertising}.
In: \bbtitle{Proceedings of the Web Conference 2021},
pp. \bfpage{2324}--\blpage{2334}
(\byear{2021})
\end{bchapter}
\endbibitem

%%% 12
\bibitem{bib12}
\begin{bchapter}
\bauthor{\bsnm{Yang}, \binits{M.}},
\bauthor{\bsnm{Li}, \binits{Q.}},
\bauthor{\bsnm{Qin}, \binits{Z.}},
\bauthor{\bsnm{Ye}, \binits{J.}}:
\bctitle{Hierarchical adaptive contextual bandits for resource constraint based
  recommendation}.
In: \bbtitle{Proceedings of the Web Conference 2020},
pp. \bfpage{292}--\blpage{302}
(\byear{2020})
\end{bchapter}
\endbibitem

%%% 13
\bibitem{bib13}
\begin{bchapter}
\bauthor{\bsnm{Glowacka}, \binits{D.}}:
\bctitle{Bandit algorithms in recommender systems}.
In: \bbtitle{Proceedings of the 13th ACM Conference on Recommender Systems},
pp. \bfpage{574}--\blpage{575}
(\byear{2019})
\end{bchapter}
\endbibitem

%%% 14
\bibitem{bib14}
\begin{barticle}
\bauthor{\bsnm{Schwartz}, \binits{E.M.}},
\bauthor{\bsnm{Bradlow}, \binits{E.T.}},
\bauthor{\bsnm{Fader}, \binits{P.S.}}:
\batitle{Customer acquisition via display advertising using multi-armed bandit
  experiments}.
\bjtitle{Marketing Science}
\bvolume{36}(\bissue{4}),
\bfpage{500}--\blpage{522}
(\byear{2017})
\end{barticle}
\endbibitem

%%% 15
\bibitem{bib15}
\begin{bchapter}
\bauthor{\bsnm{Glowacka}, \binits{D.}}:
\bctitle{Bandit algorithms in interactive information retrieval}.
In: \bbtitle{Proceedings of the ACM SIGIR International Conference on Theory of
  Information Retrieval},
pp. \bfpage{327}--\blpage{328}
(\byear{2017})
\end{bchapter}
\endbibitem

%%% 16
\bibitem{bib16}
\begin{bchapter}
\bauthor{\bsnm{Yao}, \binits{Q.}},
\bauthor{\bsnm{Chen}, \binits{X.}},
\bauthor{\bsnm{Kwok}, \binits{J.T.}},
\bauthor{\bsnm{Li}, \binits{Y.}},
\bauthor{\bsnm{Hsieh}, \binits{C.-J.}}:
\bctitle{Efficient neural interaction function search for collaborative
  filtering}.
In: \bbtitle{Proceedings of The Web Conference 2020},
pp. \bfpage{1660}--\blpage{1670}
(\byear{2020})
\end{bchapter}
\endbibitem

%%% 17
\bibitem{bib17}
\begin{bchapter}
\bauthor{\bsnm{Yao}, \binits{Q.}},
\bauthor{\bsnm{Xu}, \binits{J.}},
\bauthor{\bsnm{Tu}, \binits{W.-W.}},
\bauthor{\bsnm{Zhu}, \binits{Z.}}:
\bctitle{Efficient neural architecture search via proximal iterations}.
In: \bbtitle{Proceedings of the AAAI Conference on Artificial Intelligence},
vol. \bseriesno{34},
pp. \bfpage{6664}--\blpage{6671}
(\byear{2020})
\end{bchapter}
\endbibitem

%%% 18
\bibitem{bib18}
\begin{bchapter}
\bauthor{\bsnm{Agrawal}, \binits{S.}},
\bauthor{\bsnm{Goyal}, \binits{N.}}:
\bctitle{Thompson sampling for contextual bandits with linear payoffs}.
In: \bbtitle{International Conference on Machine Learning},
pp. \bfpage{127}--\blpage{135}
(\byear{2013}).
\bcomment{PMLR}
\end{bchapter}
\endbibitem

\end{thebibliography}


\end{document}